\documentclass[square]{ws-procs975x65}
\pdfoutput=1
\usepackage{graphicx}
\usepackage{color}
\usepackage{epsfig}
\usepackage{dcolumn}
\usepackage{bm}
\usepackage{bbm}
\usepackage{subfigure}
\usepackage{pxfonts}
\usepackage{mathrsfs}
\def\lsim{\mathrel{\rlap{\lower4pt\hbox{\hskip1pt$\sim$}}
    \raise1pt\hbox{$<$}}}                
\def\gsim{\mathrel{\rlap{\lower4pt\hbox{\hskip1pt$\sim$}}
    \raise1pt\hbox{$>$}}}                

\newcommand{\GeV}{\, \text{GeV}}
\newcommand{\MeV}{\, \text{MeV}}
\newcommand{\beq}{\begin{eqnarray}}
\newcommand{\eeq}{\end{eqnarray}}
\newcommand{\cl}{\, \rm C.L.}
\begin{document}
\title{Review of Minimal Flavor Constraints for Technicolor\footnote{This talk is based on \cite{Fukano:2009zm}  and given at 2009 Nagoya Global COE workshop `` Strong Coupling Gauge Theories in LHC Era'' (SCGT 09), December 8-11, 2009}}
\author{Hidenori S. Fukano\footnote{speaker, E-mail: hidenori@cp3.sdu.dk} and Francesco Sannino }
\address{CP$^{\bf 3}$-Origins, Campusvej 55, DK-5230 Odense M, Denmark.}

\begin{abstract}
We analyze the constraints on the the vacuum polarization of the standard model gauge bosons from a minimal set of  flavor observables valid for a general class of models of dynamical electroweak symmetry breaking.  We will show that the constraints have a strong impact on the  self-coupling and masses of the lightest spin-one resonances. 

\noindent
{\it Preprint number : CP3-Origins-2010-6}
\end{abstract}

\section{Introduction}

Dynamical electroweak symmetry breaking constitutes one of the best motivated extensions of the standard model (SM) of particle interactions. 
Walking dynamics for breaking the electroweak symmetry was introduced in \cite{Holdom:1981rm}. 
Studies of the dynamics of gauge theories featuring fermions transforming according to higher dimensional representations of the new gauge group has led to several phenomenological possibilities \cite{Sannino:2004qp,Dietrich:2005jn} such as (Next) Minimal Walking Technicolor (MWT) \cite{Foadi:2007ue}. The reader can find  in \cite{Sannino:2009za} a comprehensive review of the current status of the phase diagram for chiral and nonchiral gauge theories needed to construct sensible extensions of the SM featuring a new strong dynamics. In \cite{Belyaev:2008yj} it was launched a coherent program to investigate different signals of minimal models of technicolor at the Large Hadron Collider experiment at CERN.

Whatever is the dynamical extension of the SM it will, in general, modify the vacuum polarizations of the SM gauge bosons. LEP I and II data provided direct constraints on these vacuum polarizations \cite{Peskin:1991sw,Barbieri:2004qk}. In this talk, to be specific, we will assume that the vacuum polarizations are saturated by new spin-one states ( techni-vector meson and techni-axial vector meson) and show that it is possible to provide strong constraints on their self-couplings and masses for a general class of models of dynamical electroweak symmetry breaking. Our results can be readily applied to any extension of the SM featuring new heavy spin-one states. In particular it will severely limit the possibility to have very light spin-one resonances to occur at the LHC even if the underlying gauge theory has vanishing $S$-parameter. 

\section{Minimal $\Delta F=2$ Flavor Corrections from Technicolor}

Our goal is to compute the minimal contributions, i.e. coming just from the technicolor sector, for processes in which the flavor number $F$ changes by two units, i.e. $\Delta F = 2$. Here we consider  $F$ to be either the strange or the bottom number. 

 Besides the intrinsic technicolor corrections to flavor processes  one has also the corrections stemming out from extended technicolor models  \cite{Dimopoulos:1979es} which are directly responsible for providing mass to the SM fermions. In this talk, we are not attempting to provide a full theory of flavor but merely estimate the impact of  a new dynamical sector, per se, on well known flavor observables. We will, however, assume that whatever is the correct mechanism behind the generation of the mass of the SM fermions it will lead to SM type Yukawa interactions \cite{Chivukula:1987py}.  This means that we will constrain models of technicolor with extended technicolor interactions entering in the general scheme of minimal flavor violation (MFV) theories \cite{D'Ambrosio:2002ex}. 

\begin{figure}[h]
\begin{center}
\begin{tabular}{cccc}
{
\begin{minipage}{0.25\textwidth}
\includegraphics[scale=0.5]{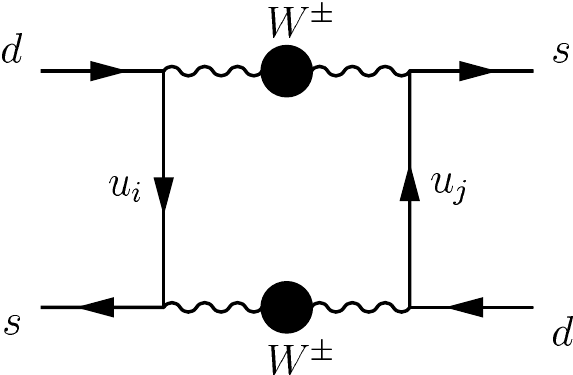}
\end{minipage}
}
{
\begin{minipage}{0.25\textwidth}
\includegraphics[scale=0.5]{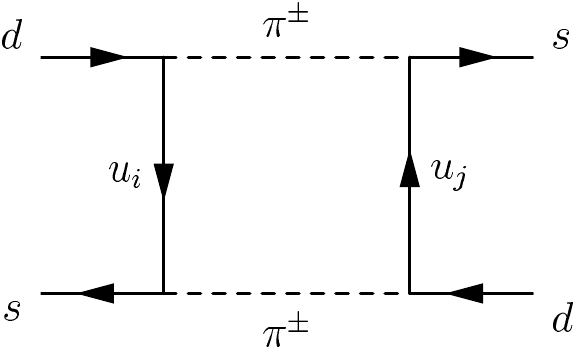}
\end{minipage} 
} 
{
\begin{minipage}{0.25\textwidth}
\includegraphics[scale=0.5]{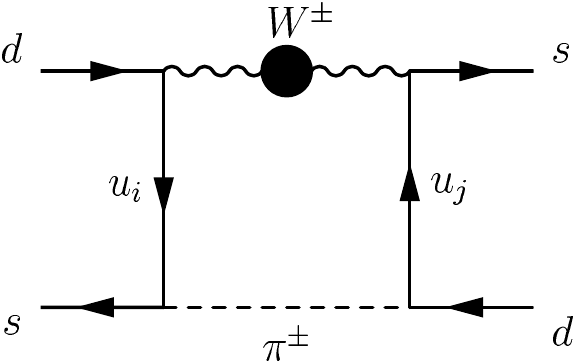}
\end{minipage}
}
{
\begin{minipage}{0.25\textwidth}
\includegraphics[scale=0.5]{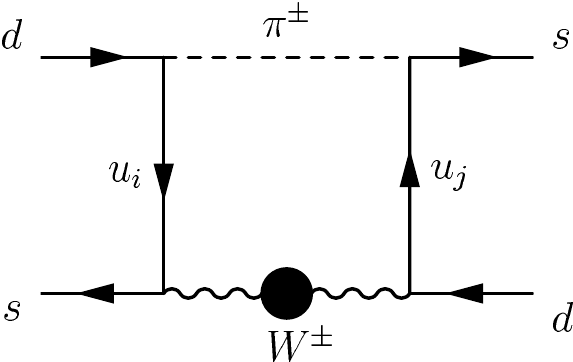}
\end{minipage} 
} 
\end{tabular}
\caption{Box diagrams for $\Delta S = 2$  {\it annihilation} processes. To obtain 
                 the $\Delta B = 2$ process, we should simply rename $s$  with $ b$ and $d$  with $ q\,(q=d,s)$ in the various diagrams.
                 \label{Box-annihilation}}
\end{center}
\end{figure}

We use the effective Lagrangian framework presented in \cite{Foadi:2007ue} according to which the relevant interactions  of the composite Higgs sector to the SM quarks up and down reads: 
\beq
{\cal L}^{\rm quark}_{\rm yukawa}
= 
\frac{\sqrt{2}\,\, m_{u_i} }{v} V_{ij} \cdot \bar{u}_{Ri} \pi^+  d_{Lj}
-
\frac{\sqrt{2}\,\, m_{d_i} }{v} V^*_{ji} \cdot \bar{d}_{Ri} \pi^- u_{Lj} 
+ h.c.\,,\label{yukawa-MWT}
\eeq 
where 
$m_{ui} , (u_i=u,c,t)$ and 
$m_{di} , (d_i = d,s,b)$ are respectively the up and down quark masses of the $i^{\rm th}$ generation. 
 $V_{ij}$ is the $i,j$ element of the Cabibbo-Kobayashi-Maskawa (CKM) matrix. 
This is our starting point which will allow us to compute the $\Delta F =2$ processes in Fig.~\ref{Box-annihilation} \footnote{
Note that the contribution of the { \it scattering} process to the invariant amplitude is equal to that of the {\it annihilation} one.}. 
After the computation of all amplitudes in Fig.~\ref{Box-annihilation}, we obtain the effective Lagrangian describing these processes as
\beq
{\cal L}^{\Delta F=2}_{\rm eff} 
=
- \frac{G^2_F M^2_W}{4 \pi^2} \cdot A(a_V,a_A)  \cdot Q_{\Delta F=2}
\,\,\,
,
\,\,\,
A(a_V,a_A)
\equiv \hspace*{-1ex}
\sum_{i,j=u,c,t} \hspace*{-1ex} \left[ \lambda_i \lambda_j \cdot E(a_i,a_j,a_V,a_A)\right]\,.
\label{lag-eff-dF=2}
\eeq
Here, we have expressed all the quantities by means of the following ratios
$ a_{\alpha} \equiv m^2_{\alpha} /M^2_W \,, (\alpha=i,j) \quad {\rm and}   \quad  a_v \equiv M^2_v/M^2_W \,, (v=V,A)$ and 
$m_i , (i=u,c,t)$ indicates the $u_i$ mass while $M_V,M_A$ are respectively the mass of the lightest techni-vector meson and techni-axial vector one. $Q_{\Delta F=2}$ and $\lambda_i$ represent 
$ \left\{ Q_{\Delta F=2} \,,\, \lambda_i \right\} = \left\{ (\bar{s}_L \gamma^\mu d_L)(\bar{s}_L \gamma_\mu d_L) \,,\, V_{id} V^*_{is} \right\}$ for $K^0-\bar{K}^0$ system 
and $ \left\{ Q_{\Delta F=2} \,,\, \lambda_i \right\} = \left\{ (\bar{b}_L \gamma^\mu q_L)(\bar{b}_L \gamma_\mu q_L) \,,\, V_{iq} V^*_{ib} \right\}$ for $B^0_q-\bar{B}^0_q$ system.
Moreover, $E(a_i,a_j,a_V,a_A)$  keeps track of the technicolor-modified gauge bosons propagators. 

Indicating with $g_{\rm EW}$ the weak-coupling constant and $\tilde{g}$ the coupling constant governing the massive spin-one self interactions and by expanding up to the order in  ${\cal O}(g^4_{\rm EW}/\tilde{g}^4)$ one can rewrite $E(a_i,a_j,a_V,a_A)$ as:
\beq
E(a_i,a_j,a_V,a_A) = E_0(a_i,a_j) + \frac{g^2_{\rm EW}}{\tilde{g}^2} \Delta E(a_i,a_j,a_V,a_A)\,.
\eeq
The SM contribution is fully contained in $E_0$ which are consistent with the results in \cite{Inami:1980fz} and the technicolor one appear first in $\Delta E$. 
The latter can be divided into a vector and an axial-vector contribution as follows:
$ \Delta E (a_i,a_j,a_V,a_A) = h(a_i,a_j,a_V) + (1 - \chi)^2 \cdot h(a_i,a_j,a_A) $
where the quantity $\chi$ was introduced first in \cite{Appelquist:1998xf} and  the axial-vector decay constant is  directly proportional to the quantity $(1-\chi)^2$. The vector and axial decay constant are: $f_V^2 = {M^2_V}/{\tilde{g}^2} \,, \, f_A^2 =  (1-\chi)^2 {M^2_A}/{\tilde{g}^2}$.

We also write: 
\beq
A(a_V,a_A)
=
A_0 + \frac{g^2_{\rm EW}}{\tilde{g}^2} \cdot \Delta A(a_V,a_A)\,.
\label{fin-A}
\eeq
Upon taking into account the unitarity of the CKM matrix and setting $a_u \to 0$ one has
$ A_0 = \eta_1 \cdot \lambda^2_c \cdot  \bar{E}_0(a_c)  + \eta_2 \cdot \lambda^2_t \cdot \bar{E}_0(a_t) + \eta_3 \cdot 2 \lambda_c \lambda_t \cdot \bar{E}_0(a_c,a_t) $ and 
$ \Delta A(a_V,a_A) = \eta_1 \cdot \lambda^2_c \cdot  \Delta \bar{E}(a_c,a_V,a_A) + \eta_2 \cdot \lambda^2_t \cdot \Delta \bar{E}(a_t,a_V,a_A) + \eta_3 \cdot 2 \lambda_c \lambda_t \cdot \Delta \bar{E}(a_c,a_t,a_V,a_A) $
where $\eta_{1,2,3}$ are the QCD corrections to $\bar{E}_0$ and $\Delta \bar{E}$. 
The explicit expressions for the functions $E_0$, $\Delta E$ , $h$, $\bar{E}$ and $\Delta \bar{E}$  various expressions can be found in \cite{Fukano:2009zm}.

We recall that the absolute value of the CP-violation parameter in the $K^0-\bar{K}^0$ system is given by~\cite{Buchalla:1995vs}:
\beq
\left( |\epsilon_K| \right)_{\rm full}
=
\frac{G^2_F M^2_W}{12 \sqrt{2} \pi^2} \times \left[ \frac{M_K}{\Delta M_K} \right]_{\rm exp.} \hspace*{-3ex} 
\times B_K f^2_K \times [ - {\rm Im}A(a_V,a_A) ] \,,
\label{def-epsilon-K}
\eeq
and the meson mass difference in the $Q^0-\bar{Q}^0\, ,\, Q=(K,B_d,B_s)$ system is given by
\beq
\left( \Delta M_Q \right)_{\rm full} 
&\equiv& 
2 \cdot \left| \langle \bar{Q}^0 | - {\cal L}^{\Delta F=2}_{\rm eff} | Q^0 \rangle \right|
=
\frac{G^2_F M^2_W}{6\pi^2} \cdot f^2_Q \cdot M_Q \times B_Q\times \left| A(a_V,a_A) \right|
\label{def-massdiff-Q}\,,
\eeq
where $f_Q$ is the decay constant of the $Q$-meson and $M_Q$ is its mass. $B_Q$ is identified with the QCD bag parameter. 
This bag parameter is an intrinsic QCD contribution and we assume that the technicolor sector does not contribute to the bag parameter~\footnote{This is a particularly good approximation when the technicolor sector does not have techiquarks charged under ordinary color. The best examples are Minimal Walking Technicolor models.  }. 
The experimental values of $G_F,M_W,f_Q,M_Q,\Delta M_Q$ and the bag parameter $B_Q$ are~\cite{Amsler:2008zzb} 
$G_F=1.1664 \times 10^{-5} \GeV^{-2}$, 
$M_W = 80.398 \GeV$, 
$f_K = 155.5 \MeV$, 
$B_K = 0.72 \pm 0.040$, 
$f_{B_d}\sqrt{B_{B_d}} = 225 \pm 35 \MeV$,
$f_{B_s} \sqrt{B_{B_s}} = 270 \pm 45 \MeV$,
$M_K = 497.61 \pm 0.02 \MeV$,
$M_{B_d} = 5279.5 \pm 0.3 \MeV$, 
$M_{B_s} = 5366.3 \pm 0.6 \MeV$.

It is convenient to define the following quantities:
\beq
\delta_\epsilon 
\equiv 
\frac{g^2_{\rm EW}}{\tilde{g}^2} \cdot 
\frac{{\rm Im}\Delta A(a_V,a_A)}{{\rm Im}A_0} \quad , \quad
\delta_{M_Q} 
\equiv 
\frac{g^2_{\rm EW}}{\tilde{g}^2} \cdot 
\frac{\Delta A(a_V,a_A)}{A_0}\,.
\label{def-mass-difference}
\eeq
Using these expressions we write $(|\epsilon_K|)_{\rm full}$ and $(\Delta M_Q)_{\rm full}$ as
$ \left( |\epsilon_K| \right)_{\rm full}= \left( |\epsilon_K| \right)_{\rm SM} \times (1 + \delta_\epsilon) \, ,\, \left( \Delta M_Q \right)_{\rm full}= \left( \Delta M_Q \right)_{\rm SM} \times \left| 1 + \delta_{M_Q} \right| $ where $\left( |\epsilon_K| \right)_{\rm SM}$ and $\left( \Delta M_Q \right)_{\rm SM}$ are the SM expressions corresponding to the representations in Eqs.(\ref{def-epsilon-K},\ref{def-massdiff-Q}) with $A = A_0$ and their values are 
$
\left( |\epsilon_K| \right)_{\rm SM} = (2.08^{+0.14}_{-0.13}) \times 10^{-3}\,, 
\left( \Delta M_K \right)_{\rm SM}= (3.55^{+1.09}_{-1.00})  \,\,\, {\rm ns}^{-1} \,,
\left( \Delta M_{B_d} \right)_{\rm SM}= (0.56^{+0.19}_{-0.16})  \,\,\, {\rm ps}^{-1}\,,\,
\left( \Delta M_{B_s} \right)_{\rm SM}= (17.67^{+6.38}_{-5.40}) \,\,\, {\rm ps}^{-1}\,
$.
The experimental values are \cite{Amsler:2008zzb}
$|\epsilon_K| = (2.229 \pm 0.012) \times 10^{-3}$,  
$\Delta M_K = 5.292 \pm 0.0009 {\rm ns}^{-1}$,
$\Delta M_{B_d} = 0.507 \pm 0.005 {\rm ps}^{-1}$, 
$\Delta M_{B_s} = 17.77 \pm 0.10 {\rm ps}^{-1}$.

We are now ready to compare the SM values with the experimental one. 
We can read off the constrain on $\delta_\epsilon$ which is:
\beq
\delta_\epsilon  =  \left( 7.05^{+ 7.93}_{-7.07} \right) \times 10^{-2} \quad (68\%\cl) \,.\label{constraint-MWT-eK}
\eeq
In order to compare the corrections associate to the kaon mass $\Delta M_K$ we formally separate the short distance contribution from the long distance one and write $ \Delta M_K = \left( \Delta M_K \right)_{\rm SD} + \left( \Delta M_K \right)_{\rm LD} $. 
Here $\left( \Delta M_K \right)_{\rm SD}$ encodes the short distance contribution which must be confronted with the technicolor one 
Eq.(\ref{def-mass-difference}) and  the long distance contribution, $\left( \Delta M_K \right)_{\rm LD}$, 
corresponds to the exchange of the light pseudoscalar mesons. It is difficult to pin-point the $\left( \Delta M_K \right)_{\rm LD}$ contribution ~\cite{Buchalla:1995vs,Donoghue:1992dd} and hence we can only derive very weak constraints  from $\delta_{M_K}$. In fact we simply require that $ \left( \Delta M_K \right)_{\rm SD} = \left( \Delta M_K \right)_{\rm SM} \,  |1 + \delta_{M_K}| 
\leq \left( \Delta M_K \right)_{\rm exp.} $. This means that:
\beq
|1 + \delta_{M_K}|  \leq 2.08 \quad (68\%\cl) \,. \label{constraint-MWT-dMK}
\eeq
On the other hand  the short distance contribution dominates the $B^0_q-\bar{B}^0_q$ mass difference~\cite{Donoghue:1992dd} yielding the following constraints:
\beq
 |1 + \delta_{M_{Bd}}|  = 0.91^{+0.38}_{-0.24} \quad , \quad
 |1 + \delta_{M_{Bs}}|  = 1.01^{+0.44}_{-0.27} \qquad (68\%\cl) \,. \label{constraint-MWT-dMB}  
\eeq

\section{Constraining Models of Dynamical Electroweak Symmetry Breaking}
We will now use the minimal flavor experimental information to reduce the parameter space of a general class of models of dynamical electroweak symmetry breaking.

If the underlying technicolor theory is QCD like we can impose the standard 1st Weinberg sum rule (WSR) : $ f^2_V- f^2_A = f^2_\pi = \left( {v_{\rm EW}}/{\sqrt{2}}\right)^2$ and 2nd WSR : $f^2_V \, M^2_V - f^2_A \, M^2_A = 0$ 
with $f_V$ and $f_A$ the vector and axial decay constants as shown in  \cite{Foadi:2007ue,Appelquist:1998xf,Duan:2000dy}. 
Using the explicit expressions of the decay constants in terms of the coupling $\tilde{g}$ and vector masses provided in \cite{Appelquist:1998xf,Duan:2000dy,Foadi:2007ue} and imposing the above sum rules we derive:
$ 1/a_V = {(g^2_{\rm EW} S)}/{(16\pi)} -  {1}/{a_A} \,\, ( \geq 0 ) $, with the $S$-parameter~\cite{Peskin:1991sw} reading \cite{Appelquist:1998xf,Duan:2000dy,Foadi:2007ue}: 
$S \equiv 8\pi \left[ {f^2_V}/{M^2_V} - {f^2_A}/{M^2_A} \right] = \left( {8 \pi} /{\tilde{g}^2} \right) \left[ 1 - (1 - \chi)^2 \right]$ . 
The condition above yields the following additional constraint for $\tilde{g}$ by simply noting that the quantity $(1-\chi)^2$  is positive: 
$\tilde{g} < \sqrt{{8\pi}/{S}} $. The constraints on $(M_A, \tilde{g})$ induced by WSRs and theoretical constraints are stronger, for a given $S$, than the ones deriving from flavor experiments and expressed in (\ref{constraint-MWT-eK})-(\ref{constraint-MWT-dMB}). This is not surprising given that in an ordinary technicolor theory the spin-one states are very heavy.

The situation changes when allowing for a walking behavior. Besides the flavor constraints one has also the ones due to the electroweak precision measurements  \cite{Foadi:2007se} as well as the unitarity constraint of $W_L-W_L$ scattering \cite{Foadi:2008xj}. We will consider all of them. As for the walking technicolor case we reduce the number of independent parameters at the effective Lagrangian level via the 1st WSR : $ f^2_V- f^2_A = f^2_\pi = \left( {v_{\rm EW}}/{\sqrt{2}}\right)^2$ and the 2nd modified \cite {Appelquist:1998xf}  WSR : $f^2_V \, M^2_V - f^2_A \, M^2_A = a \cdot ({16 \pi^2} f^4_\pi)/d({\rm R})$ where $a$ is a number expected to be positive and ${\cal O}(1)$~\cite{Appelquist:1998xf} and  $d({\rm R})$ is the dimension of the representation of the underlying technifermions as shown in \cite{Foadi:2007ue}.  We have now $ M^2_A < {8\pi f^2_\pi}/{S} \cdot (2-\chi)/(1-\chi) = {8\pi f^2_\pi}/{S} \cdot  \left[ 1 + ( 1-{\tilde{g}^2 S}/{(8\pi)})^{-1/2}\right] $.

\begin{figure}[h!]
\begin{center}
\includegraphics[width=0.3\textwidth,height=0.3\textwidth]{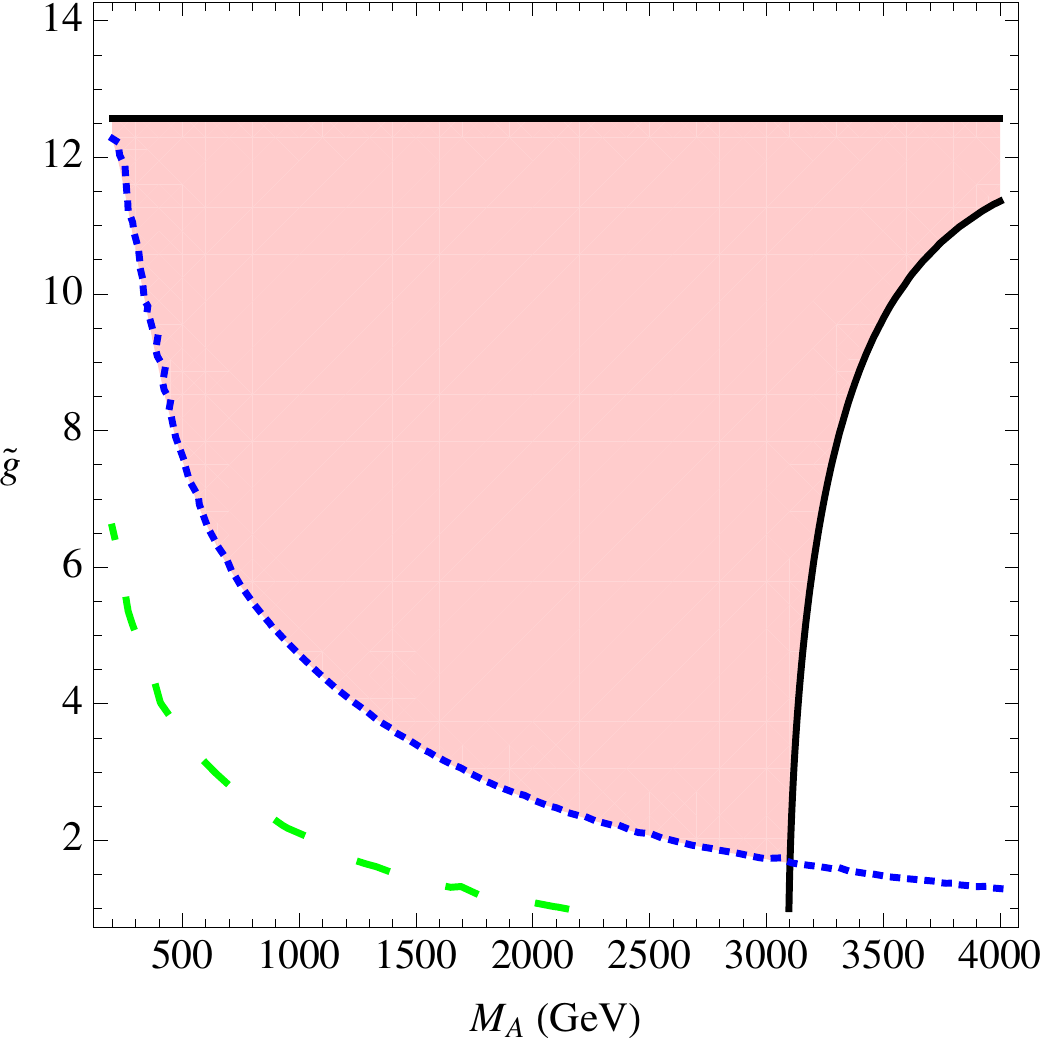} \hspace*{5ex}
\includegraphics[width=0.3\textwidth,height=0.3\textwidth]{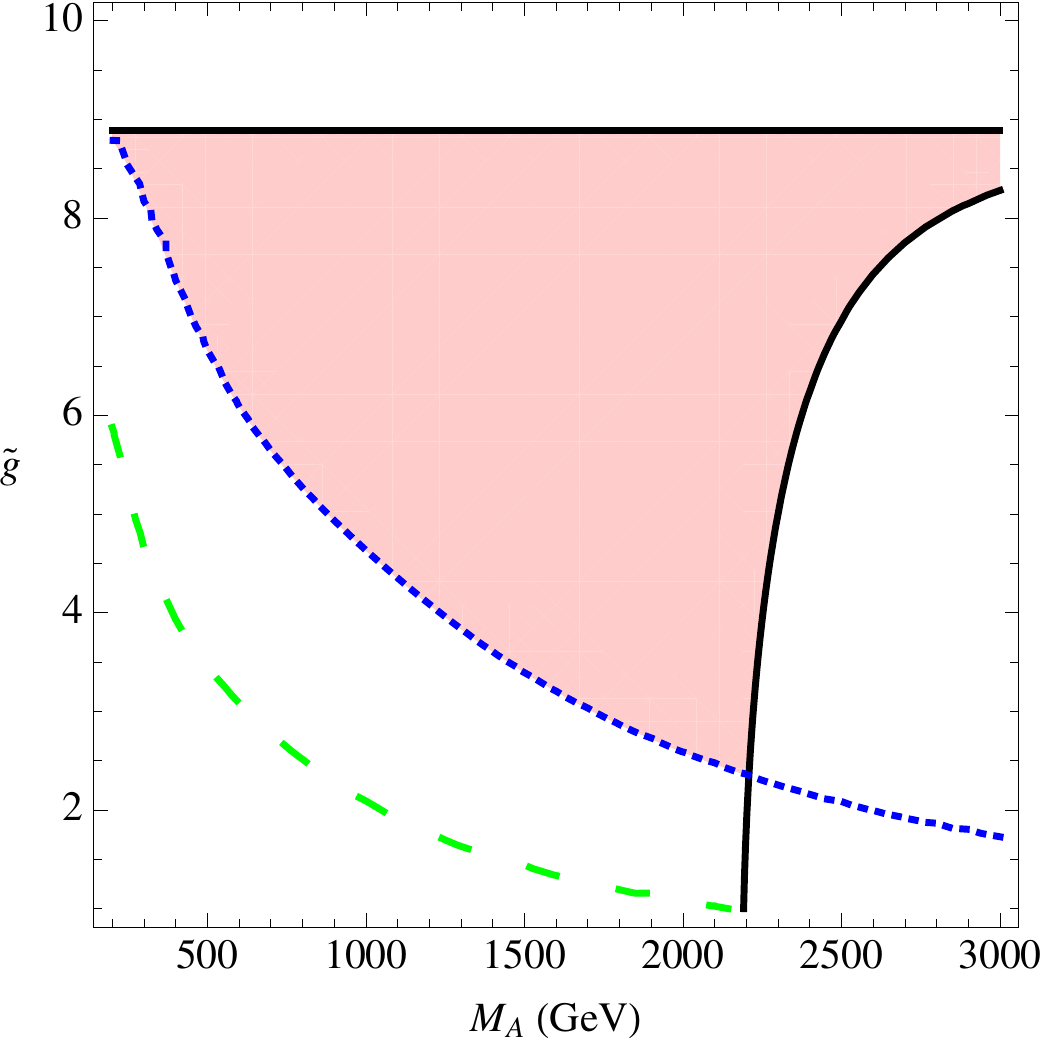}\\[2ex]
\includegraphics[width=0.3\textwidth,height=0.3\textwidth]{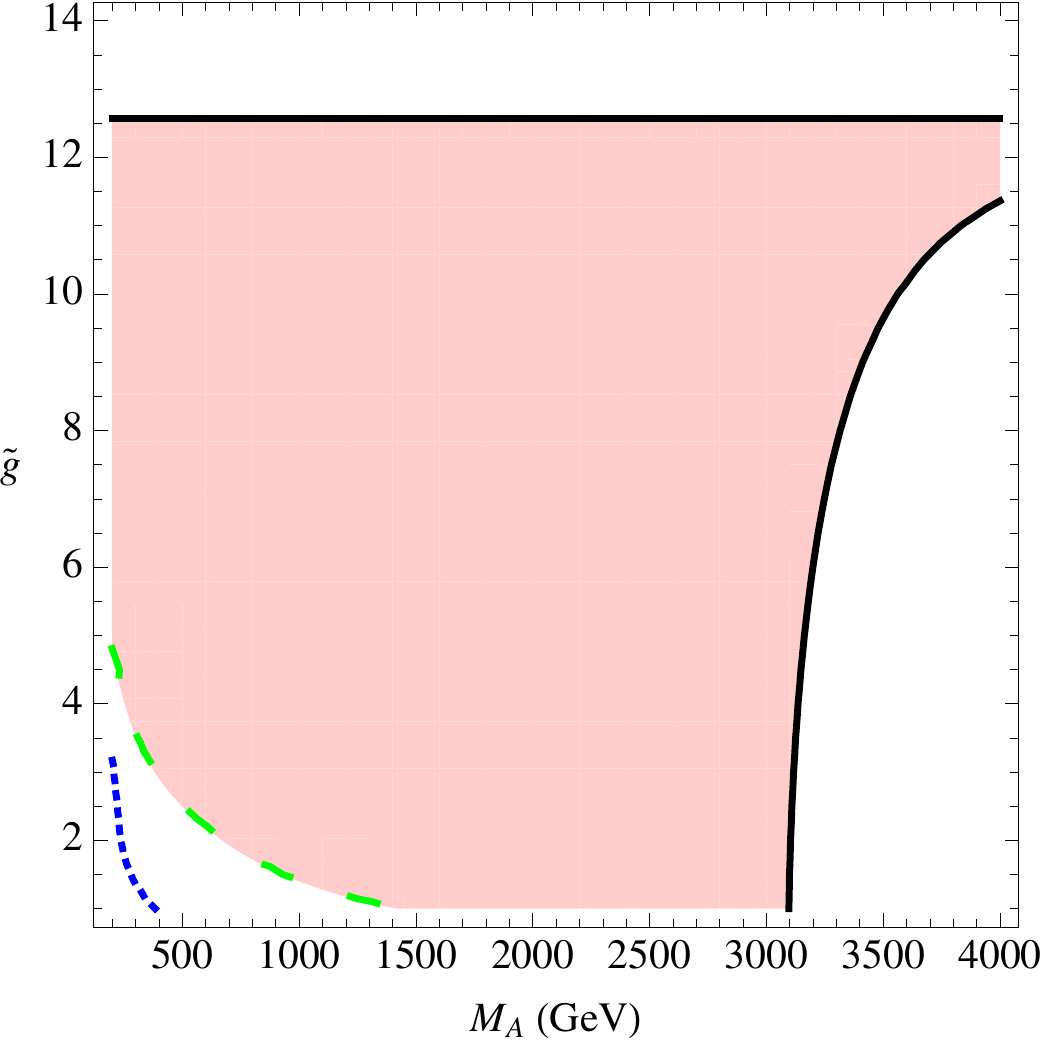} \hspace*{5ex}
\includegraphics[width=0.3\textwidth,height=0.3\textwidth]{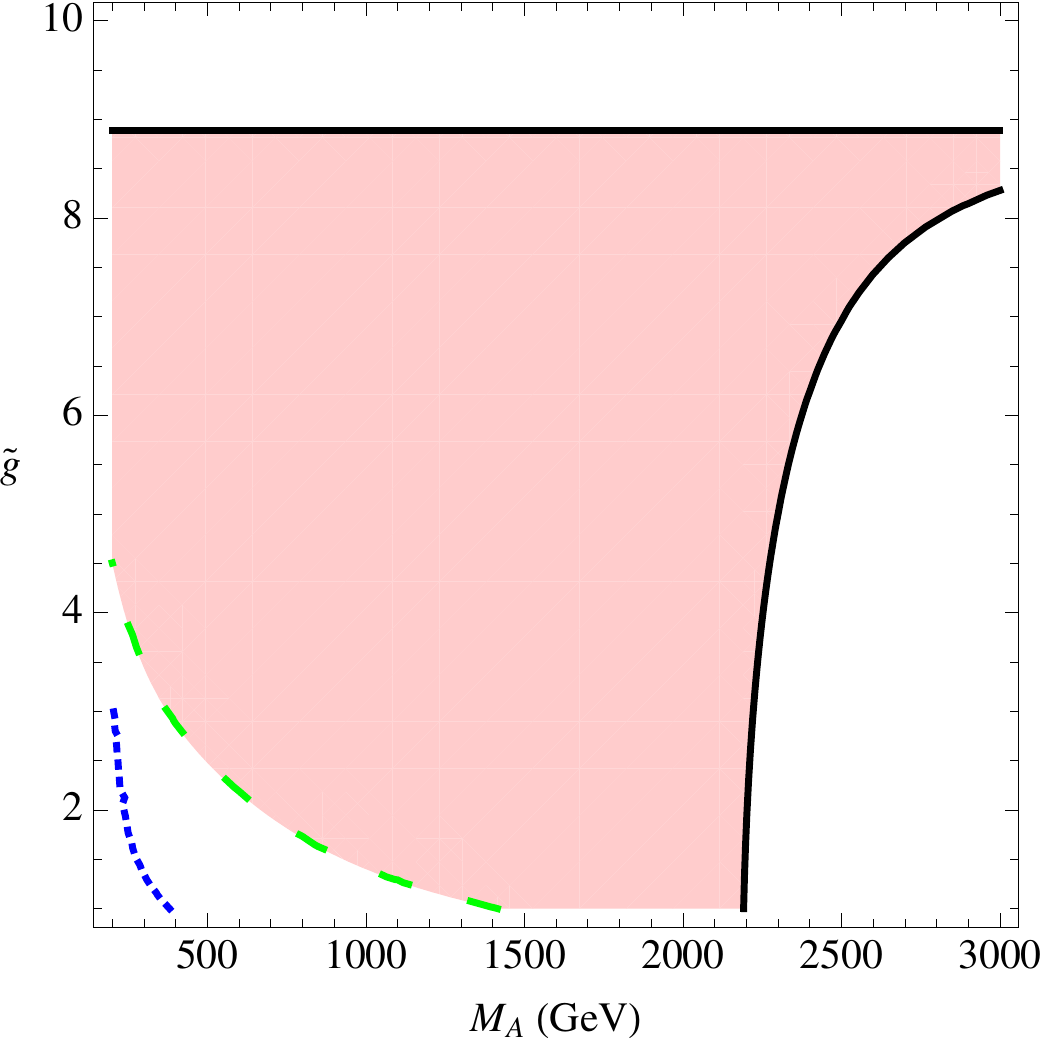}
\caption{The upper and lower left panels represents the allowed region in the $(M_A,\tilde{g})$-plane for MWT respectively for the $68\%\cl$ and  $95\%\cl$. A similar analysis is shown for NMWT in the right hand upper and lower panels. 
The region above the straight solid line is forbidden by the condition $\tilde{g} < 12.5$  for MWT and  $\tilde{g} < 8.89$ for NMWT while the region below the solid  curve (on the right corner) is forbidden by theoretical upper bound for $M_A$.
In the two upper (lower) plots the dotted lines correspond to the $68\%\cl$ ($95\%\cl$) flavor constraints while the dashed lines are the $68\%\cl$ ($95\%\cl$)   from LEP II data. 
The flavor constraints come only from $\epsilon_K$ since the ones from $\Delta M_{B_q}$ are not as strong.  }                  
\label{plot-constraints-walking}
\end{center}
\end{figure}

In Fig.~\ref{plot-constraints-walking}, we show the allowed region in the $(M_A,\tilde{g})$-plane after having imposed the minimal flavor constraints  due to the experimental values of  $|\epsilon_K|$ and $\Delta M_Q$  obtained using Eqs.~(\ref{constraint-MWT-eK})-(\ref{constraint-MWT-dMB}) together with the theoretical constraints for $\tilde{g},M^2_A$.
To obtain Fig.~\ref{plot-constraints-walking}, we used the expressions for $A_0$ and $\Delta A(a_V,a_A)$ in which $a_V = \left[ 1- \tilde{g}^2 S /(8\pi ) \right] \cdot a_A + 2 \tilde{g}^2/g^2_{\rm EW}$. Given that the upper bound for $\delta_{M_K}$ is always larger than the theoretical estimate in the region $M_A>200\GeV$ we conclude that the  $\Delta M_K$ constraint is not yet very severe and hence it is not displayed in Fig.~\ref{plot-constraints-walking}. To make the plots we need also the value of the $S$ parameters and hence we analyzed as explicit example minimal walking technicolor models. In the case of MWT,  we use for $S$ the naive MWT estimate,  i.e. $S=1/(2\pi)$~\cite{Sannino:2004qp} while $\tilde{g}$ is constrained to be $\tilde{g} < 12.5$. In the  case of Next to Minimal Walking Technicolor (NMWT), we use the the naive $S$ is approximately $1/\pi$~\cite{Sannino:2004qp} and the constraint on $\tilde{g}$ yields  $\tilde{g} < 8.89$. 

We have plotted the various constraints on the $(M_A,\tilde{g})$-plane for MWT(NMWT) in the upper and lower left (right) panel of Fig.~\ref{plot-constraints-walking}.  In the upper  (lower) left figure we compare the $68\%\cl$  ($95\%\cl$) allowed regions coming from the minimal flavor constraints (the darker region above the dotted line) with the ones from LEP II data (region above the dashed line).  It is clear that the flavor constraints are stronger for the $68\%\cl$ case but are weaker for the $95\%\cl$ one with respect to the constraints from LEP II data. 

\begin{figure}[h!]
\begin{center}
\includegraphics[width=0.3\textwidth,height=0.3\textwidth]{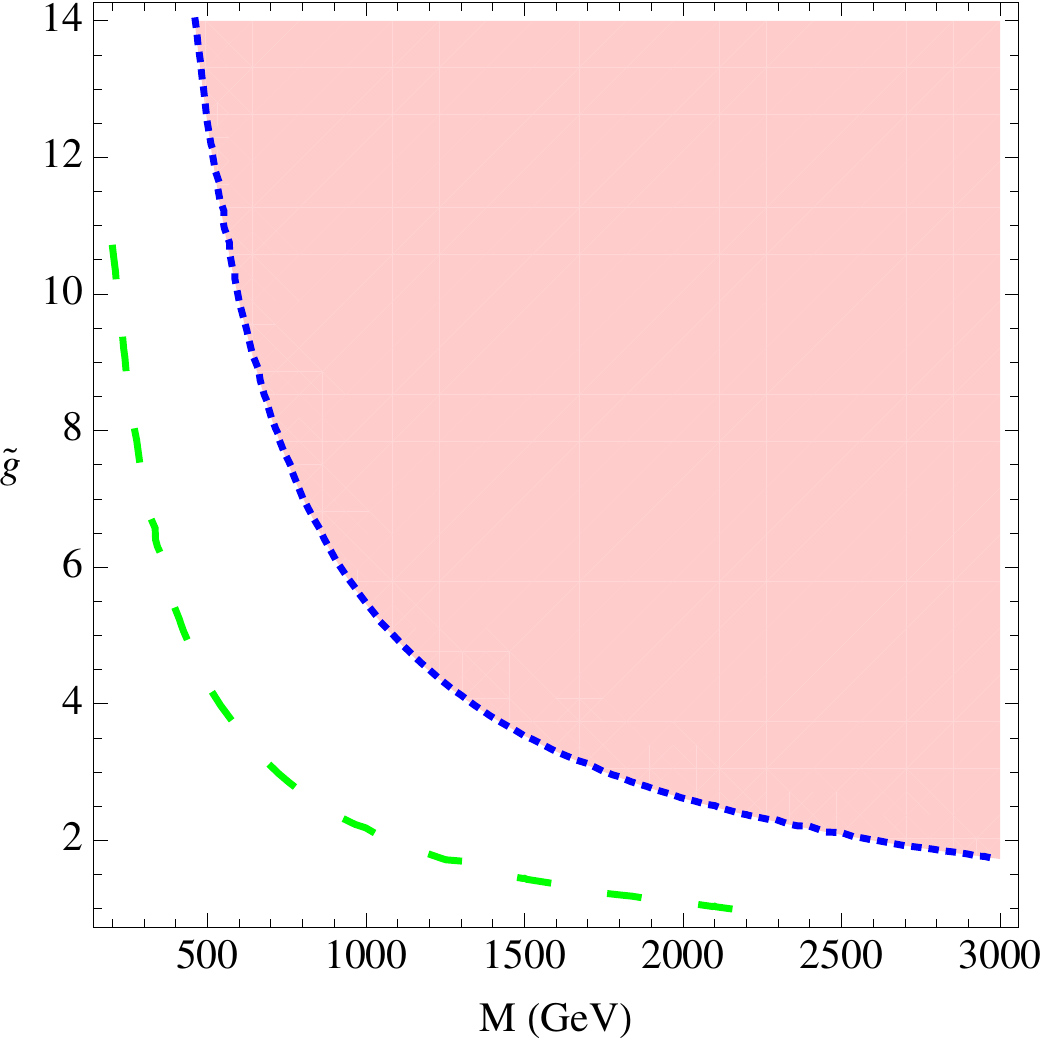} \hspace*{5ex}
\includegraphics[width=0.3\textwidth,height=0.3\textwidth]{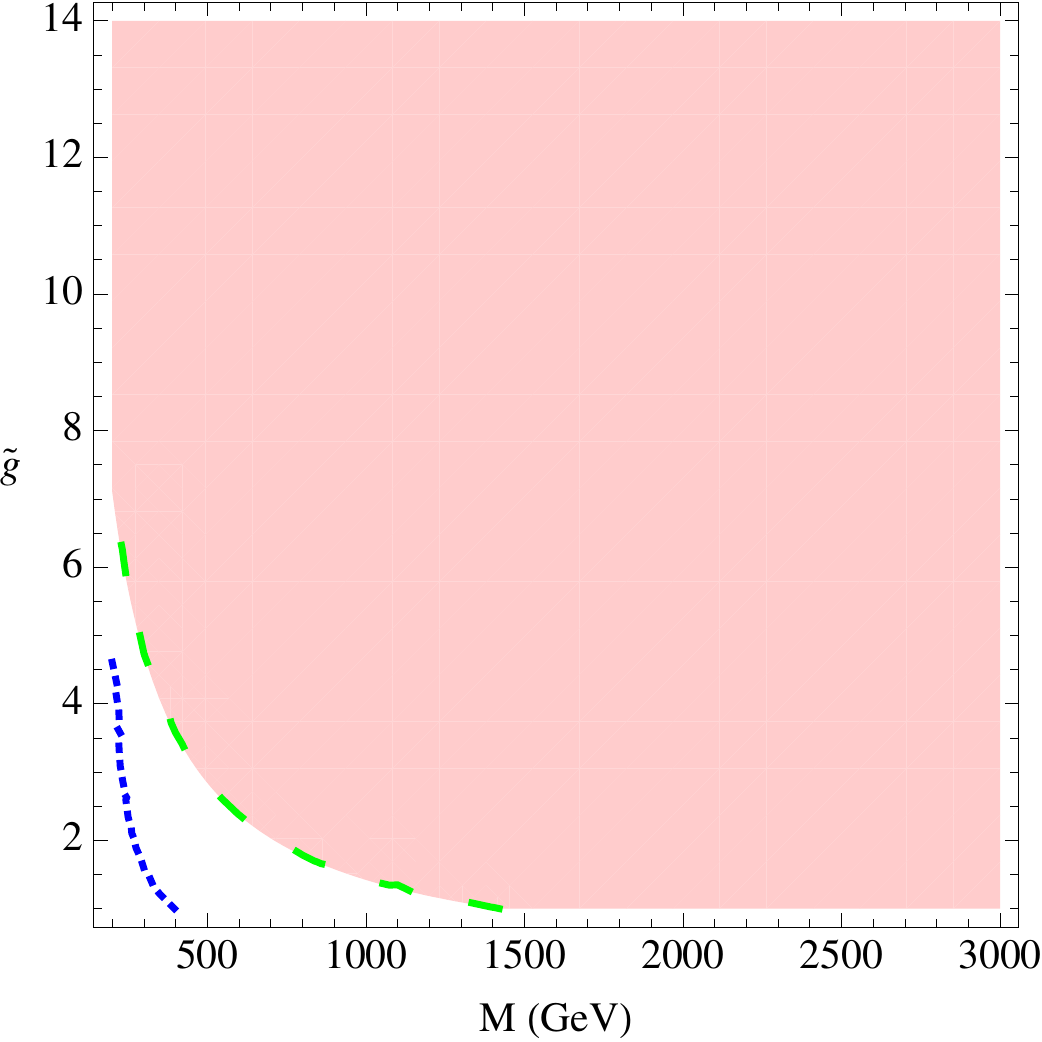}
\caption{The left (right) panel represents the allowed region in the $(M_A,\tilde{g})$-plane for CT respectively for the $68\%\cl$ ($95\%\cl$). 
In the two upper (lower) plots the dotted lines correspond to the $68\%\cl$ ($95\%\cl$) flavor constraints while the dashed lines are the $68\%\cl$ ($95\%\cl$)   from LEP II data. 
The flavor constraints come only from $\epsilon_K$ since the ones from $\Delta M_{B_q}$ are not as strong.  }                  
\label{plot-contraints-CT}
\end{center}
\end{figure}

In the limit $M_A= M_V = M$ and $\chi =0$ the effective theory acquires a new symmetry  \cite{Appelquist:1999dq}. This new symmetry relates a vector and an axial field and can be shown to work as a custodial symmetry for the $S$ parameter \cite{Appelquist:1999dq}. The only non-zero electroweak parameters are $W,Y$ parameters. It was already noted in \cite{Foadi:2007se} that a custodial technicolor model cannot be easily achieved via an underlying walking dynamics and should be interpreted as an independent framework. This is so since custodial technicolor models do not respect the WSRs \footnote{One can, of course, imagine more complicated vector spectrum leading to such a symmetry.}. We directly compare in the Fig.~\ref{plot-contraints-CT} the constraints on the custodial technicolor parameter region ($M$,$\tilde{g}$) coming from LEP II and flavor constraints and find a similar trend as for the other cases. 

 \section{Summary}

Flavor constraints are  relevant for models of dynamical electroweak symmetry breaking  with light spin-one resonances, in fact,  any model featuring spin-one resonances with the same quantum numbers of the SM gauge bosons will have to be confronted with these flavor constraints. It would be interesting to combine the present analysis with the one presented in \cite{Antola:2009wq} in which a low energy effective theory for the ETC sector was introduced.


\end{document}